\documentclass[%
 reprint,
 amsmath,amssymb,
 aps,
]{revtex4-1}

\usepackage{graphicx}
\usepackage{dcolumn}
\usepackage{bm}
\usepackage[pagebackref=false,colorlinks,linkcolor=black,citecolor=black]{hyperref}

\begin{document}

\preprint{APS/123-QED}

\title{Generalized entropies and the expansion law of the universe}

\author{Fatemeh Lalehgani Dezaki}
 \altaffiliation[f.lalehgani@ph.iut.ac.ir ]{ Department of Physics, Isfahan University of Technology, Isfahan, 84156-83111, Iran.}
\author{  Behrouz Mirza}%
 \email{b.mirza@cc.iut.ac.ir}
\affiliation{%
 Department of Physics, Isfahan University of Technology, Isfahan, 84156-83111, Iran.
}%




\begin{abstract}
We suggest that using the first law of thermodynamics is a convenient method to obtain a correct form of the expansion law of the universe \cite{T. Padmanabhan1}. We will, then, use this idea to obtain the expansion law for a Kodama observer. By using the expansion law for a Kodama observer, we can obtain the dynamic equation of the FRW universe for deformed Horava-Lifshitz gravity. The use of the first law of thermodynamics also leads to a new approach for obtaining  the Friedmann equations  for f(R) and scalar tensor gravities.

\keywords{\textbf{Modified gravities, Horava- Lifshitz gravity, F(R) gravity, The first law of thermodynamics }}
\end{abstract}

\pacs{Valid PACS appear here}
\maketitle


\section{\label{sec:level1}Introduction}
In 1995, Jacobson \cite{T. Jacobson} derived Einstein equations by demanding fundamental Clausius relations $\delta Q=TdS$ for all the local Rindler horizon through each spacetime point, with $\delta Q$ and T interpreted as the energy flux and Uunruh temperature seen by the accelerated observer. In the FRW universe, it is shown  \cite{R.-G. Cai} that the Friedmann equation which describes the dynamics of the universe can be derived from the first law of thermodynamics associated with apparent horizon by assuming a temperature $T=\frac{1}{2\pi \tilde{r}_A} $ and an entropy $S=\frac{A}{4G}$, where $\tilde{r}_A$ and A are the radius and area of the apparent horizon, respectively. Recently, it has been proposed that the expansion of the universe is attributed to the difference between the number of degrees of freedom on a holographic surface and the number of freedom in the bulk. In this paradigm, the dynamic equation of a FRW universe can be successfully derived. It has also been argued  that for a pure de sitter universe with the Hubble constant H, the holographic principle can be expressed as $N_{surf} = N_{bulk}$, where $ N_{surf} $ denotes the number of  degrees of freedom on the spherical surface with the Hubble radius $\frac{1}{H}$ for a flat FRW universe and $N_{bulk}$ is the bulk degrees of freedom obeying the equipartition law, $N_{bulk} = \frac{2|E|}{T}$. Our universe is not exactly de sitter but there is considerable evidence that it is asymptotically de sitter. This would suggest that the emergence of space should be related with $ (N_{sur} -N_{bulk}) $ \cite{T. Padmanabhan1}.\\
An improved expansion law has been suggested  for a non-flat FRW universe in \cite{A. Sheykhi}.
However, the origin of the expansion law is not yet well understood and there is not a general method to obtain the expansion law for a flat or non-flat universe.  In this paper, attempts will be made to show that the first law of thermodynamics can be considered as the origin of the expansion law of the universe, or alternatively, that we can exploit the first law of thermodynamic as a convenient method to obtain the expansion law of the universe. We can derive the expansion law of the universe for a dynamic Kodama observer. In this case, we use the unified first law of thermodynamics to obtain a correct form of the expansion law of the universe. A new gravity theory at a Lifshitz point proposed by Horava that referred as the Horava-Lifshitz theory. In this gravity due to the thermal fluctuations and quantum fluctuations, we should apply the  quantum corrections to the horizon area \cite{Horava P,Horava1}. For f(R) and the scalar tensor gravity, due to the non-equilibrium thermodynamics of spacetime, an additional entropy production term has to be added to the Clausius relation \cite{M. Akbar1}, while this is not the case  for  Einstein gravity. Unlike Einstein gravity, we should consider a generalized Misner-Sharp energy in these modified gravities \cite{Misner1,R. M. Wald,L. D. Landau,Sharp,Rong-Gen Cai}. The dynamic equation of FRW universe for the Horava- Lifshitz gravity and Friedmann equations for f(R) and scalar tensor gravities can be obtained based on the above considerations.\\
It is known that Einstein's field equations can be obtained from the first law of thermodynamics, so we may represent the relationship between  the first law of thermodynamics, Einstein's field equations, and the expansion law of the universe on a triangle as in Fig. \ref{figure 1}. It is interesting that by using  only one of these relations, the two others can be derived.\begin{figure}[h]
\centering
\includegraphics[width=0.45\textwidth]{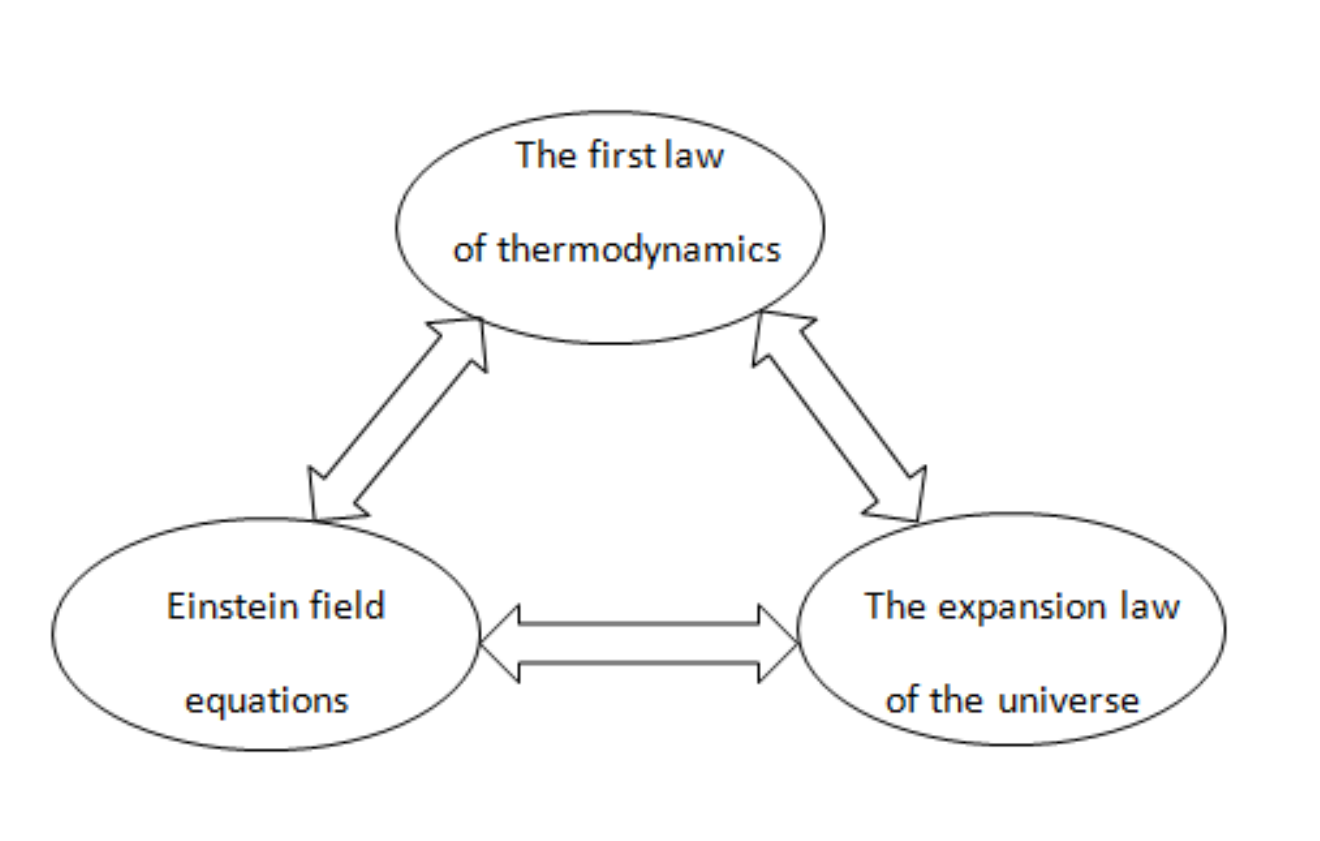}
\caption[]{\it{The first law of thermodynamics and its relationships with the expansion law of the universe and the Einstein field  equations. }} \label{fig:figure 1}
\end{figure}
The rest of this paper is organized as follows. In Section 2, we will review and illustrate the first law of thermodynamics as the origin of the expansion law of the universe. In Section 3,  the unified first law of thermodynamics will be used to obtain the the expansion law of the universe for a dynamic Kodama observer and we use the expansion law for Kodama observer to obtain the dynamic equation of FRW universe for deformed Horava-Lifshitz gravity. In Section 4, we will investigate f(R) and scalar tensor gravities  using our approach.

\section{\label{sec:level2}REVIEW OF THE FIRST LAW OF THERMODYNAMICS AND THE EXPANSION LAW OF THE UNIVERSE}

In this part, we are going to review and  explain how to derive the expansion law of the universe from  the first law of thermodynamics so that we may interpret the first law of thermodynamic as the origin of the expansion law of the universe.

 It is a rather difficult task to define energy for a gravitational field in general relativity. Due to the absence of the local energy density of gravitational field, it is tempting to define some meaningful quasi-local energy so that  on the boundary of a given region in spacetime. Indeed, it is possible to properly define such quasi-local energies as the Misner-Sharp energy \cite{Misner1,Sharp}. Various properties of the Misner-Sharp energy are discussed in some detail by Hayward in \cite{Hayward1,Hayward2}. It was shown that the Misner Sharp energy surrounded by the apparent horizon $\tilde{r}_{A}$ of the FRW universe is nothing but the total energy $\rho V$ of matter inside the apparent horizon where, $\rho$ is the matter density and $V$ is the volume of the sphere with radius $\tilde{r}_A=\frac{1}{\sqrt{H^2+\frac{k}{a^2}}}\nonumber$.  $H=\frac{\dot{a}}{a}$ is the Hubble parameter and k is the  spatial curvature of the universe. On the apparent horizon of a FRW universe, the energy crossing the apparent horizon within the time interval $dt$ is given by:
\begin{eqnarray}\label{form-2}
\delta Q &=& dE_{MS}|_{\tilde{r}_A}\nonumber \\
&=& A(\rho + P) H\tilde{r}_A dt\nonumber \\
&=&3H(\rho + P)Vdt \\
&=& T dS \nonumber
\end{eqnarray}
where, $P$ is the matter pressure, $A$ is the surface of the sphere with radius $\tilde{r}_A $, and $T$ is the Hawking temperature $T=\frac{1}{2\pi \tilde{r}_A}$ \cite {Ya-Peng Hu}. It should be noted that we have  used an approximate form of the unified first law of thermodynamics $(dE=TdS+WdV)$ i.e. we assumed that the apparent horizon radius is constant $(\frac{\dot{\tilde{r}}_A}{2H\tilde{r}_A}\ll 1)$ and so $T=\frac{1}{2\pi \tilde{r}_A}$ and the first law of thermodynamic can be written as $dE=TdS$, for more details see \cite {R.-G. Cai,Qiao- Jun Cao}. Therefore,
\begin{equation}\label{form-4}
 \frac{dE_{MS}}{dt}=T\frac{dS}{dt}
 \end{equation}
Using Equations (\ref{form-2}) and $E_{MS}=\rho V$, the above equation can be rewritten as follows
\begin{equation}\label{form-5}
\frac{dE_{MS}}{dt}=H((\rho + 3P)  V+ 2E_{MS})
\end{equation}

Now, we want to obtain the expansion law for a non-flat FRW universe by using the first law of thermodynamics. At first step, we have to define the bulk degrees of freedom.  Padmanabhan  \cite {T. Padmanabhan1,T. Padmanabhan2} suggested  that the bulk degrees of freedom $N_{bulk}$  is related to the Komar energy \cite {Arture Komar,T. Padmanabhan} by $N_{bulk}=\frac{2 |E_{Komar}|}{T}=\frac{2|(\rho + 3P)V|}{T}$. It should be noted that we cannot prove this relation and it is just postulated in \cite {T. Padmanabhan1}. We also assume the same relation for definition of the bulk degrees of freedom. Now by replacing $(\rho + 3P)V=-\frac{N_{bulk}T}{2}$ in equation  (\ref{form-5}) we obtain

\begin{eqnarray}\label{form-63}
 (\frac {-N_{bulk}}{2} )T+ 2E_{MS} = \frac{1}{H} \frac{dE_{MS}}{dt}
\end{eqnarray}


It is interesting that the second term in Equation (\ref{form-63}) can be  related to the surface degrees of freedom ($N_{sur}=\frac{A}{L_{P}^2}$). To illustrate, we use Equation (\ref{form-4}) to obtain:


\begin{eqnarray}\label{form-10}
E_{MS} &=& T \int dS\nonumber\\
&=& T (\frac{A}{4L_{P}^2})\\
&=& \frac{1}{4}T N_{sur}
\end{eqnarray}
Therefore,
\begin{equation}\label{form-11}
(\frac{ 2}{T}) 2E_{MS} = N_{sur}
\end{equation}
Using Eq. (\ref{form-4}), the  last term in Equation (\ref{form-63}) leads to an increase  in  volume
 \begin{eqnarray}\label{form-12}
\frac{2}{T} \frac{1}{H}\frac{dE_{MS}}{dt}&=&(\frac{2}{TH}) \frac{1}{2\pi\tilde{r}_A}\frac{d}{dt}(\frac{A}{4 L _{p}^2})\nonumber\\
&=&\frac{1}{L _{p}^2 H \tilde {r}_A}\frac{dV}{dt}
\end{eqnarray}
(The relation holding between  volume V and  area A with a radius $\tilde{r}_A$ is $ \frac{dV}{dA} = \frac{\tilde{r}_A}{2}$.) Therefore,
it is easily seen that the expansion law of  universe, can be derived as follows

\begin{equation}\label{form-13}
\frac{dV}{dt} = L_{p}^2H\tilde{r}_A(N_{sur} - N_{bulk})
\end{equation}

\noindent Thus, we have shown that the expansion law of the universe can be obtained from the first law of thermodynamics.  Equation  (\ref{form-13}) has already been obtained in \cite{A. Sheykhi} by considering $T=\frac{1}{2\pi \tilde{r}_A}$ i.e. when the changing in apparent horizon is infinitesimal.  In the next section, we are going to obtain the expansion law of the universe for a dynamic Kodama observer by using the unified first law of thermodynamics $(dE=TdS+WdV)$. Also we will obtain the modified dynamic equation of FRW universe in a Horava-Lifshits gravity.

\section{\label{sec:level3}Modified dynamic equation of FRW universe in a Horava-
Lifshitz gravity for a Kodama observer}
There is no timelike killing vector in the dynamic spacetime. The Kodama vector generates a preferred flow of time and is a dynamic analogue of a stationary Killing vector \cite{Kodama}. For a Kodama observer, we use the unified first law of thermodynamics and the Hawking temperature  detected by a Kodama observer and defined as follows \cite{S. A. Hayward5,Cai3,Frolov,U. H. Daniels-Son,R. Bousso,G. Calcagni}
\begin{eqnarray}\label{form-14}
T&=&-\frac{1}{2\pi \tilde{r}_A}(1-\frac{\dot{\tilde{r}}_A}{2H\tilde{r}_A})\nonumber\\
&=&-\frac{1}{2\pi\tilde{r}_A}(1-\epsilon)
\end{eqnarray}
where,
\begin{equation}\label{form-15}
\epsilon=\frac{\dot{\tilde{r}}_A}{2H\tilde{r}_A}
\end{equation}
We assume that $\epsilon<1$. In other words, we are dealing with an "inner" trapping horizon  rather than an "outer" trapping one (with a positive temperature).

Now, we investigate the expansion law for the asymptotic de sitter universe in view of the unified first law of thermodynamics. The unified first law of thermodynamics is expressed as follows  \cite{S. A. Hayward5,Cai3}
\begin{equation}\label{form-16}
dE_{MS}=TdS+WdV
\end{equation}
where,
\begin{eqnarray}\label{form-17}
W &=&\frac{ (\rho-P)}{2}\nonumber\\
V&=&\frac{4}{3}\pi \tilde{r_A}^3
\end{eqnarray}
The energy crossing the apparent horizon within the interval $dt$ for this unified first law is \cite{M-Akbar2}
\begin{equation}\label{form-18}
dE_{MS} =- AH \tilde{r}_A ((1- 2 \epsilon) \rho + P) dt
\end{equation}
And
\begin{equation}\label{form-19}
T dS =- (1 - \epsilon) AH \tilde{r}_A ( \rho + P) dt
\end{equation}
\begin{equation}\label{form-20}
W dV =  \epsilon AH \tilde{r}_A ( \rho - P ) dt
\end{equation}
Therefore, the unified first law of thermodynamics can be written as follows
\begin{equation}\label{form-21}
\frac{dE_{MS}}{dt}=T \frac {dS}{dt}+W\frac{dV}{dt}
\end{equation}
Using Equations (\ref{form-19}) and (\ref{form-20}), the above equation can be rewritten as follows
\begin{equation}\label{form-22}
\frac{dE_{MS}}{dt}=-H((\rho + 3P)V +2\rho V- 6\epsilon \rho V)
\end{equation}
Therefore, by considering  the Misner-Sharp energy, $E_{MS}=\rho V$, we have
\begin{equation}\label{form-23}
-\frac{1}{H}\frac{dE_{MS}}{dt}=(\rho + 3P)V+(2-6 \epsilon) E_{MS}
\end{equation}
Using (\ref{form-23}), the expansion law of universe for Kodama observer can be obtained.

Reorganizing Equation (\ref{form-23})  with the help of Equations (\ref{form-14}),(\ref{form-18}), and (\ref{form-19}), we have
\begin{eqnarray}\label{form-24}
(\rho + 3P)V&=&-\frac{1}{H} \frac{dE_{MS}}{dt}-2E_{MS}(1-3\epsilon)\nonumber\\
&=& -\frac{T}{H( 1- \epsilon)} \frac{dS}{dt} -2E_{MS} \nonumber\\
&=& +\frac{1}{2 \pi {\tilde{r}}_AH}\frac{dS}{dt} -2E_{MS}
\end{eqnarray}
The Misner-Sharp energy for the first law of thermodynamics can be obtained as follows \cite {Yi- Xin Chan}
\begin{eqnarray}\label{form-25}
E_{MS}&=&2TS+3WV\nonumber\\
&=&2T(\frac{A}{4L^2_{P}})+3WV
\end{eqnarray}
With the help of Equation (\ref{form-17}) and considering $(\rho + 3P)V=\frac{T}{2}N_{bulk}$ we get (\ref{form-26}) below:
\begin{eqnarray}\label{form-26}
2E_{MS}&=&(\rho + 3P)V - TN_{sur}\nonumber\\
&=&\frac{T}{2}N_{bulk}-TN_{sur}
\end{eqnarray}
The above relation can be rewritten as the expansion law of the universe for a dynamic Kodama observer as follows
\begin{equation}\label{form-30}
\frac{dV}{dt} = -TA H {L_{p}^2} (N_{sur} - N_{bulk})
\end{equation}
Thus, it is shown that to obtain the expansion law for a Kodama observer, we have to use  the unified first law of thermodynamics. We, therefore, suggest the application of this fundamental relation of thermodynamics as the correct method  for obtaining the expansion law and its generalized versions. For another approach see \cite{Mostafa}. It should be noted that equation (\ref{form-30}) is obtained for a Kodama observer where the Hawking temperature is defined in Eq. (\ref{form-14}).   We cannot obtain (\ref{form-13}) just by replacing $T$ by  $1/2 \pi \tilde{r}_A$ in (\ref{form-30}). For obtaining equation (\ref{form-13}), we have to put $\epsilon=0$ in (\ref{form-14}), (\ref{form-19}) and (\ref{form-20}),  and then do the rest of calculations. It is  interesting  that different observers use similar but distinct forms of the expansion law of the universe. We may argue that origin of this difference may go back to different forms of the first law of thermodynamics for different observers. In the following we will obtain modified dynamic equation of FRW universe in a deformed Horava-Lifshitz gravity  for a Kodama observer.
One of the quantum corrections applied to the horizon area is logarithmic correction that arises from loop quantum gravity due to thermal equilibrium fluctuations and quantum fluctuations. As a result, the entropy for deformed Horava-Lifshitz gravity is written as \cite{Li-Ming,A. Castillo}.
\begin{equation}
S=\frac{A}{4L_{p}^2}+\frac{\pi}{\omega} \ln(\frac{A}{L_{p}^2})
\end{equation}
where $\omega$ is a dimensionless constant parameter of the theory. The number of degrees of freedom on the apparent horizon $(N=4S)$ can be written as follows
\begin{equation}
N_{sur}=\frac{A}{L_{p}^2}+\frac{4 \pi}{\omega} \ln(\frac{A}{L_{p}^2})
\end{equation}
For a Kodama observer the Hawking temperature defined as follows
\begin{equation}
T=-\frac{1}{2\pi \tilde{r}_A}(1-\frac{\dot{\tilde{r}}_A}{2H\tilde{r}_A})
\end{equation}
By considering $\frac{dV_{eff}}{dt}=2L_{p}^2 \tilde{r}_{A} \frac{dS_{eff}}{dt}$ \cite{caii} we can arrive to following relation
\begin{eqnarray}
\frac{dV_{eff}}{dt}&=&  4 \pi \tilde{r}^2_{A}  \dot{\tilde{r}}_A + \frac{4 \pi L^2_{P} \dot{\tilde{r}}_A}{ \omega}\nonumber\\
&=&\frac{-4\pi (\dot{H}-\frac{k}{a^2})}{(H^2 + \frac{k}{a^2})^2}-\frac{4 \pi L^2_{P} (\dot{H}-\frac{k}{a^2})}{(H^2 + \frac{k}{a^2})\omega}
\end{eqnarray}
Substituting the above relations in Equation (\ref{form-30}) we find the dynamic equation of FRW universe for deformed Horava-Lifshitz gravity
\begin{eqnarray}
H^2 +\dot{H}&=&-\frac{4 \pi L^2_{P}}{3}(\rho+ 3 P)+\frac{L^2_{P}(H^2 + \frac{k}{a^2})^2}{\omega}\nonumber\\
&\times &\ln(\frac{4\pi}{L^2_{P} (H^2 + \frac{k}{a^2})})
\nonumber\\
&~~~+&\frac{L^2_{P} (H^2 + \frac{k}{a^2})(\dot{H}-\frac{k}{a^2})}{2\omega}\nonumber\\
&\times &\Big(1+\ln(\frac{4\pi}{L^2_{P} (H^2 + \frac{k}{a^2})})\Big)
\end{eqnarray}
As expected, setting $\omega \rightarrow \propto$, the standard dynamic equation of FRW universe can be obtained.

\section{\label{sec:level4}FRIEDMANN EQUATION FOR F(R) AND SCALAR TENSOR GRAVITIES}
 The Hawking temperature together with the black hole entropy is connected through the identity $TdS = dE$ and is usually called the first law of black hole thermodynamics. It should be noted that the usual Clausius relation $\delta Q = TdS$ holds for neither f(R) nor scalar tensor gravities. Hence, an internal entropy needs to be produced  to balance the energy conservation, $\delta Q = TdS + Td\bar{S}$ with \cite {M. Akbar1}
\begin{equation}\label{form-31}
d\bar{S}=\frac{-A}{4L_{p}^2}(H{\tilde{r}_A}^2(d(f''(R)\dot{R})-Hf''(R)\dot{R})+f''(R)\dot{R})dt
\end{equation}
where, $d(f''(R)\dot{R})\equiv d(f''(R)\dot{R})/dt$ and $S=Af'(R)/4L_{p}^2$ is the horizon entropy of the black hole.
We further define ${S_{eff}}=S+\bar{S}$ as an effective entropy associated with the apparent horizon of the FRW universe. So, the field equation of f(R) gravity can be rewritten as $ d(E_{MS})_{eff} = Td{S_{eff}}$, where $T=\frac{1}{2\pi \tilde{r}_A}$.\\ Unlike  Einstein gravity, the existence of the generalized Misner-Sharp energy depends on a constraint  that should be satisfied in f(R) gravity. For f(R) gravity, the generalized Misner-Sharp energy $(E_{Ms})_{eff}$ can be expressed as \cite {Rong-Gen Cai}
\begin{eqnarray}\label{form-32}
(E_{MS})_{eff}&=&\frac{\tilde{r}_A^3}{2{L_{p}^2}}(\frac{1}{{\tilde{r}_A}^2} f'(R)+\frac{1}{6}(f(R)\nonumber\\
&-&f'(R)R)+Hf''(R)\dot{R})
\end{eqnarray}

\noindent where, $\frac{dV_{eff}}{dt}$ is related to $\frac{dS_{eff}}{dt}=\frac{dS}{dt}+\frac{d\bar{S}}{dt}$ as follows
\begin{eqnarray}\label{form-33}
\frac{d{V_{eff}}}{dt}&=&\frac{\tilde{r}_A}{2}\frac{d{A_{eff}}}{dt}\nonumber \\
&=& 2L_{p}^2\tilde{r}_A\frac{d{S_{eff}}}{dt}\nonumber \\
&=& 2\pi \tilde{r}_A\frac{d}{dt}(r^2f'(R))-2\pi \tilde{r}_A^3(H\tilde{r}_A^2(d(f''(R)\dot{R})\nonumber\\
&~~~-&Hf''(R)\dot{R})+f''(R)\dot{R})
\end{eqnarray}
 Using Equations (10) and (32), we obtain the effective area as follows
\begin{eqnarray}\label{form-34}
{A_{eff}}&=&8\pi {L_{p}^2} \tilde{r}_A (E_{MS})_{eff}\nonumber \\
&=&4\pi \tilde{r}_A^4(\frac{1}{\tilde{r}_A^2}f'(R)\nonumber\\
&~~~+&\frac{1}{6}(f(R)-f'(R)R)+Hf''(R)\dot{R})
\end{eqnarray}

The effective number of degrees of freedom on the apparent horizon is $ N_{sur}=\frac{{A_{eff}}}{L_{p}^2}$ and the cosmic volume is $V=\frac{4\pi {\tilde{r}_A}^3}{3}$. By considering  $(\rho + 3P)V=\frac{T}{2}N_{bulk}$, we  can also obtain the bulk degrees of freedom
\begin{equation}\label{form-35}
N_{bulk}={\frac{-16 {\pi} ^2{\tilde{r}_A}^4}{3}(\rho +3P)}
\end{equation}
Substituting the above relations into (\ref{form-13}), we obtain the dynamic equation for f(R) gravity
\begin{eqnarray}\label{form-36}
H^2+\dot{H}&=&\frac{-4\pi {L_{p}^2}}{3}(\frac{\rho +3P}{f'(R)}\nonumber\\
&~~~+&\frac{\rho _{curv} + 3P_{curv}}{8\pi {L_{p}^2}})
\end{eqnarray}
where,
\begin{equation}\label{form-37}
\rho_{curv}=\frac{1}{f'}\Big(\frac{-(f(R)-Rf'(R))}{2}-3H f''(R)\dot{R}\Big)
\end{equation}

\begin{eqnarray}\label{form-38}
P_{curv}&=&\frac{1}{f'}\Big(\frac{(f(R)-Rf'(R))}{2}+f''(R) \stackrel{..}{R}\nonumber\\
&~~~+&f'''(R) \dot{R}^2+2Hf''(R)\dot{R}\Big)
\end{eqnarray}

Integrating (\ref{form-36}) and using the continuity equation, $\dot{\rho}_{total} + 3H (\rho_{total} + P_{total})=0$, we  get the familiar form of  Freidmann equations of the Friedmann-Robertson-Walker universe for f(R) gravity.

\begin{equation}
H^2+\frac{k}{a^2}= \frac{8\pi L_{p}^2}{3}\Big (\frac{\rho }{f'}+\frac{{\rho} _{curv}}{8\pi L_{p}^2}\Big)
\end{equation}

\begin{equation}\label{form-39}
\dot{H}-\frac{k}{a^2}=-4\pi {L_{p}^2}\Big (\frac{(\rho +P)}{f'}+\frac{({\rho} _{curv}+P_{curv})}{8\pi L_{p}^2}\Big)
\end{equation}
The above Friedmann equations was obtained from the expansion law of the universe.

For a different approach to f(R) gravity we will adopt the assumption $\tilde {S}=\frac{A f'(R)}{4G}$, without the term that relates to the non-equilibrium thermodynamics of spacetime, therefore $\frac{d{\tilde V}}{dt}$ is as follows
\begin{eqnarray}
\frac {d{\tilde V}}{dt}&=&\frac{\tilde{r}_A}{2}\frac{d{\tilde{A}}}{dt}\nonumber \\
&=& 2L_{p}^2\tilde{r}_A\frac{d{\tilde{S}}}{dt}\nonumber \\
&=& 2\pi \tilde{r}_A\frac{d}{dt}(r^2f'(R))
\end{eqnarray}
The number of degrees of freedom on the apparent horizon for this entropy  is $N_{sur}=\frac{\tilde{A}}{L_{p}^2}$

where
\begin{equation}
\tilde{A}=(4\pi \tilde{r}_A^2) f'(R)
\end{equation}
In this way the effective terms relating to the non-equilibrium thermodynamics can be obtain by the energy-momentum tensor, Therefore the Komar energy is as follows
\begin{eqnarray}
E_{Komar}&=&\int_{V} dV(2T_{\mu\nu}^{(t)} - T^{(t)} g_{\mu\nu}) u^{\mu} u^{\nu}\nonumber\\
&=&V(T_{00}^{(t)}+T_{ii}^{(t)})\nonumber\\
&=&V\Big((\rho +\tilde{\rho})+3(P +\tilde{P})\Big)
\end{eqnarray}
where
\begin{equation}
\tilde \rho =\frac{-(f(R)-Rf'(R))}{2}-3H f''(R)\dot{R}
\end{equation}

\begin{eqnarray}
\tilde P&=&\frac{(f(R)-Rf'(R))}{2}+f''(R) \stackrel{..}{R}\nonumber\\
&~~~+&f'''(R) \dot{R}^2+2Hf''(R)\dot{R}
\end{eqnarray}
Then the bulk degrees of freedon is given by
\begin{equation}
N_{bulk}={\frac{-16 {\pi} ^2{\tilde{r}_A}^4}{3}\Big( (\rho+\tilde{\rho}) +3 ( P+\tilde{P})}\Big)
\end{equation}

It should be noted that additional terms due to the non- equilibrium thermodynamics can be apear into the energy- momentom tensor.
With the above considerations we can derive the Friedmann equation for this gravity
\begin{equation}
H^2=\frac{8 \pi G}{3f'(R)}(\rho +\tilde \rho)
\end{equation}
In the first method, we obtain Friedmann equation by using generalized entropy that is equivalente to
 its effect on the energy-momentum tensor.

Also Friedmann equation can be obtained by assuming $S=\frac{A}{4G}$ \cite{Yi Ling}. In this way the additional terms can be obtaion from the Komar energy
\begin{eqnarray}
E_{Komar}&=&V\Big(\frac{1}{f'(R)}\Big(\rho +\Big (\frac{-\Big(f(R)-Rf'(R)\Big)}{2}\nonumber\\
&-&3H f''(R)\dot{R}\Big)\nonumber\\
&+& 3\frac{1}{f'(R)}\Big(P +\frac{\Big(f(R)-Rf'(R)\Big)}{2}\nonumber\\
&+&f''(R) + f'''(R) \dot{R}^2+2Hf''(R)\dot{R})\Big)
\end{eqnarray}
With the above definitions Friedmann equation can be easilly derived.

The same strategy may be applied to the scalar-tensor gravity. The entropy S of this gravity is of the form \cite{M. Akbar and Rong- Gen Cai}
\begin{equation}\label{form-40}
S=\frac{Af(\phi)}{{4L_{p}^2}}
\end{equation}
And, the Clausius relation for this gravity is given by
\begin{equation}\label{form-41}
\delta Q=TdS + Td\bar{S}
\end{equation}
where,
\begin{equation}\label{form-42}
d\bar{S}=\frac{-A}{4{L_{p}^2}}(H{\tilde{r}_A}^2(8\pi {L_{p}^2}({\dot{\phi}}^2 +\stackrel{..}{f}(\phi) -3H\dot{f}(\phi))+\dot{f}(\phi))
\end{equation}
The generalized Misner-Sharp energy $(E_{MS})_{eff}$ in this case can be explicitly expressed by
\begin{eqnarray}\label{form-43}
(E_{MS})_{eff}&=&\frac{{\tilde{r}_A}^3}{2{L_{p}^2}}\Big(\frac{1}{{\tilde{r}_A}^2}f(\phi)\nonumber\\
&~~~-& \frac{8\pi {L_{p}^2}}{3}\Big(\frac{\dot {\phi}^2}{2}+v(\phi)-3H\dot{f(\phi)}\Big)\Big)
\end{eqnarray}
Following the same procedure as the one used  for f(R) gravity, we obtain $\frac{dV_{eff}}{dt}$ that is related to $\frac{dS_{eff}}{dt}=\frac{dS}{dt}+\frac{d\bar{S}}{dt}$ for the scalar tensor gravity
\begin{eqnarray}\label{form-44}
\frac{d{V_{eff}}}{dt}&=&\frac{\tilde{r}_A}{2}\frac{d{A_{eff}}}{dt}\nonumber \\
&=& 2L_{p}^2\tilde{r}_A\frac{d{S_{eff}}}{dt}\nonumber \\
&=&2\pi \tilde{r}_A \frac{d}{dt}(r^2f(\phi))-2\pi {\tilde{r}_A}^3\Big(H{\tilde{r}_A}^2(8\pi {L_{p}^2}\Big({\dot{\phi}}^2 \nonumber\\
&~~~+&\stackrel{..}{f}(\phi) -3H\dot{ f}(\phi)\Big)+\dot{f}(\phi)\Big)
\end{eqnarray}
and the effective area is given by
\begin{eqnarray}\label{form-45}
A_{eff}&=& 8\pi {L_{P}^2} \tilde{r}_A (E_{MS})_{eff} \nonumber \\
&=&4\pi {\tilde{r}_A^4}\Big(\frac{1}{{\tilde{r}_A}^2}f(\phi)- \frac{8\pi {L_{p}^2}}{3}\Big(\frac{\dot{\phi}^2}{2}+V(\phi)\nonumber\\
&~~~-&3H\dot{f}(\phi)\Big)\Big)
\end{eqnarray}

Using the above equation, we can define the effective number of degrees of freedom on the apparent horizon.
The number of degrees of freedom in the bulk will be the same as that in (\ref{form-35}).
Substituting the above relations into the generalized expansion law (\ref{form-13}), we arrive at the dynamic equation for scalar tensor gravity for the FRW universe
\begin{equation}\label{form-46}
\dot{H}+H^2=\frac{-4\pi {L_{p}^2}}{3}\Big(\frac{(\rho +3P)}{f(\phi)}+\frac{\rho ^{(curv)} + 3P^{(curv)}}{8\pi L_{p}^2}\Big)
\end{equation}
where,
\begin{equation}\label{form-47}
\rho _{curv}=\frac{1}{f(\phi)}\Big(\frac{{\dot{\phi}}^2}{2}+V(\phi)-3H\dot{f(\phi)}\Big)
\end{equation}
\begin{equation}\label{form-48}
P_{curv}=\frac{1}{f(\phi)}\Big(\frac{{\dot{\phi}}^2}{2}-V(\phi)+2H\dot{f(\phi)}+\stackrel{..}f(\phi)\Big)
\end{equation}
Integrating on the  above equations and using the continuity equation, we can obtain the Friedmann equations as follows
\begin{equation}\label{form-49}
H^2+\frac{k}{a^2}=\frac{8\pi {L_{p}^2}}{3}\Big(\frac{ \rho}{f(\phi)}+\frac{\rho _{curv}}{8\pi L_{p}^2}\Big)
\end{equation}

\begin{equation}
\dot{H}-\frac{k}{a^2}=-4\pi {L_{p}^2}\Big (\frac{(\rho +P)}{f(\phi)}+\frac{({\rho} _{curv}+P_{curv})}{8\pi L_{p}^2}\Big)
\end{equation}
That is the familiar forms of the Friedmann equations for scalar tensor gravity.

\section{\label{sec:citeref}CONCLUSIONS}
In this paper, we suggest that using  the first law of thermodynamics is  a convenient method for obtaining  the expansion law of the universe. It was shown that we can use the unified first law of thermodynamics to obtain the expansion law for the Kodama observer. It is known that the first law of thermodynamics is equivalent to Einstein field equations. So, we expect  a relationship to exist between the first law of thermodynamics, Einstein field equations, and the expansion law of the universe.  We obtain the dynamic equation of FRW universe for the Horava-Lifshitz gravity by using the expansion law for Kodama observer. We have  also obtained the Friedmann equations for f(R) and scalar tensor gravities by this approach. Our method might be used as a general and simple approach for investigating other modified gravities.


\begin{thebibliography}{99}
\bibitem{T. Padmanabhan1}
T. Padmanabhan, Emergence and Expansion of Cosmic Space as due to the Quest for Holographic Equipartition, arXiv: 1206.4916v1[hep-th].

\bibitem{T. Jacobson}
T. Jacobson, Thermodynamics of Spacetime: The Einstein Equation of State, Phys. Rev. Lett \textbf{75}, 1260 (1995), arXiv: [gr-qc/9504004].

\bibitem{R.-G. Cai}
R.-G. Cai and S. P. Kim, First law of thermodynamics and Friedmann equatins of Friedmann- Robertson- Walkeruniverse, JHEP \textbf{02},  (2005) 050, arXiv: [hep-th/0501055]

\bibitem{A. Sheykhi}
A. Sheykhi, Friedmann equations from emergence of cosmic space, Phys. Rev. D \textbf{87}, 061501 (R) (2013), arXiv: 1304.3054 [gr-qc].

\bibitem{Horava P}
P. Horava,  Spectral dimension of the universe in Quantum Gravity at a Lifshitz point, Phys. Rev. Lett, \textbf{102}, 161301, (2009), arXiv: 0902.3657 [hep-th].

\bibitem{Horava1}
P. Horava, On asymtotic Darkness in the Horava- Lifshitz Gravity, Phys. Rev. D, \textbf{79}, 084008, (2009), arXiv:1110.0036 [hep-th].

\bibitem{M. Akbar1}
M. Akbar, Rong-Gen Cai, Thermodynamic Behavior of Field Equations for f(R) Gravity, Phys.Lett. B \textbf{648}, 243-248, (2007), arXiv: [gr-qc/0612089].

\bibitem{Misner1}
C. W. Misner, K. S. Thorne and J. A. Wheeler, Gravitation (Freeman, San Francisco 1973).

\bibitem{R. M. Wald}
 R. M. Wald, General Relativity (Chicago, The university of chicago Press, 1984).

 \bibitem{L. D. Landau}
 L. D. Landau and E. M. Lifshitz, The classical Theory of Field (Beijing, Butterworth-Heinmann, 1999).

\bibitem{Sharp}
C. W. Misner and D. H. Sharp, Spherical gravitational collapse with energy transport by radiative diffusion, Phys. Rev. D 136, B 571 (1964).

\bibitem{Rong-Gen Cai}
Rong-Gen Cai, Li-Ming Cao, Ya-Peng Hu, Nobuyoshi Ohta, Generalized Misner-Sharp Energy in f(R) Gravity, Phys. Rev. D \textbf{80}, 104016, (2009), arXiv: 0910.2387 [hep-th].

\bibitem{Hayward1}
S. A. Hayward, Gravitational Energy in Spherical Symmetry, Phys. Rev. D 53 (1996) 1938-1949.

\bibitem{Hayward2}
S. A. Hayward, Quasilocal gravitational energy, Phys. Rev. D \textbf{49}, 831 (1994).

\bibitem{Ya-Peng Hu}
Rong-Gen Cai, Li-Ming Cao, Ya-Peng Hu, Hawking Radiation of Apparent Horizon in a FRW Universe, Class. Quant. Grav.  \textbf{26}, 155018, (2009), arXiv: 0809.1554 [astro-ph.CO].

\bibitem{Qiao- Jun Cao}
Qiao- Jun Cao, Yi- Xin Chen, Kai- Nan Shao, Clausius relation and Friedmann equation in FRW universe model, JCAP 05 (2010) 030, arXiv: 1001.2597 [hep-th].

\bibitem{T. Padmanabhan2}
T. Padmanabhan, Emergent perspective of Gravity and Dark Energy, arXiv: 1207.0505 [astro-ph.CO].

\bibitem{Arture Komar}
Arture Komar, Covariant Conservation Law In General Relativity, Phys. Rev. D  \textbf{113}, 934 (1959).

\bibitem{T. Padmanabhan}
T. Padmanabhan, Phys. Rev. D \textbf{81}, 124040, (2010), arXiv: 1003.5665 [hep-th].

\bibitem{Kodama}
H. Kodama, Progr. Theor. Phys.  \textbf{63}, 1217 (1980).

\bibitem{S. A. Hayward5}
S. A. Hayward, unified first law of black-hole dynamics and relativistic thermodynamics, Class. Quant. Grav \textbf{15} (1998), arXiv: [gr-qc/9710089].

\bibitem{Cai3}
Rong-Gen Cai,  Li-Ming Cao, Unified First Law and Thermodynamics of Apparent Horizon FRW Universe, Phys. Rev. D \textbf{75}, 064008, (2007), arXiv: [gr-qc/0611071].

\bibitem{Frolov}
A. V. Frolov and L. Kofman, Inflation and de Sitter Thermodynamics, JCAP \textbf{0305}, (2003) 009, arXiv: [hep-th/0212327].

\bibitem{U. H. Daniels-Son}
 U. H. Danielsson, Transplanckian energy production and slow roll inflation, Phys. Rev. D 71 (2005) 023516, arXiv: [hep-th/ 0411172].

\bibitem{R. Bousso}
 R. Bousso, Cosmology and the S-matrix, Phys. Rev. D 71 (2005) 064024, arXiv: [hep-th/0412197].

 \bibitem{G. Calcagni}
 G. Calcagni, de Sitter thermodynamics and the braneworld, JHEP 0509, (2005) 060, arXiv: [hep-th/0507125].

\bibitem{M-Akbar2}
M. Akbar, Rong-Gen Cai, Thermodynamic behavior of the Friedmann equation at the apparent horizon of the FRW universe ,Phys. Rev. D \textbf{75}, 084003, (2007), arXiv: [hep-th/0609128].

\bibitem{Yi- Xin Chan}
Yi- Xin Chan, Jian- Long Li, Yang- Qiang Wang, Thermodynamics for Kodama observer in general spherically symmetric spactime, arXiv: 1008.3215 [hep-th].

\bibitem{Mostafa}
M. Hashemi, S. Jalalzadeh and S. Vasheghani Farahani, Hawking temperature and the emergent cosmic space, arXiv: 1308.2383[gr-qc].

 \bibitem{Li-Ming}
Rong-Gen Cai, Li-Ming Cao, Nobyyoshi Ohta, Topological Black Holes in Horava-Lifshitz Gravity, Phys. Rev. D \textbf{80}, 024003, (2009), arXiv: 0904.3670 [hep-th].

\bibitem {A. Castillo}
A. Castillo and A. Larranga, Entropy for black holes in the deformed Horava- Lifshitz gravity, EJTP \textbf{8}, 25 (2011), arXiv: 0906.4380[gr-qc].

 \bibitem {caii}
Rong- Gen Cai, Emergence of Space and Spacetime Dynamics of Friedmann- Robertson- Walker, JHEP 11 (2012) 016, arXiv: 1207.0622 [gr-qc].

\bibitem{Yi Ling}
Yi Ling, Wen-Jian Pan, Note on the emergence of cosmic space in modified gravities, Phys. Rev. D \textbf{88}, 043518, (2013), arXiv: 1304.0220[hep-th].

\bibitem{M. Akbar and Rong- Gen Cai}
M. Akbar and Rong- Gen Cai, Friedmann equations of FRW universe in scalar-tensor gravity, f(R) gravity and the first law of  thermodynamics, B 635 : 7-10 (2006) [arXiv: hep-th/0602156].
\end{thebibliography}
\end{document}